\documentstyle[psfig]{aa}

\newcommand{\beq}{\begin{equation}}
\newcommand{\eeq}{\end{equation}}
\newcommand{\noi}{\noindent}
\newcommand{\pks}{ PKS 0439$-$433 }

\newcommand{\dV}{\Delta V}

\newcommand{\Msun}{$M_\odot$}
\newcommand{\MHI}{$M_{\rm H~{\sc i}}$}

\begin{document}

\thesaurus{12(12.03.3), 11(11.09.1 PKS~0439-433), 13(13.19.3)}

\title{ATCA search for 21~cm emission from a candidate damped Ly-$\alpha$ absorber at $z = 0.101$.}

\titlerunning{ATCA search for 21~cm emission at $z = 0.101$}

\author{Nissim Kanekar \inst{1}\thanks{nissim@ncra.tifr.res.in}, 
Jayaram N Chengalur\inst{1,4}\thanks{chengalu@ncra.tifr.res.in}, 
Ravi Subrahmanyan\inst{2}\thanks{rsubrahm@atnf.csiro.au} \& 
Patrick Petitjean\inst{3}\thanks{ppetitje@iap.fr}}

\authorrunning{Nissim Kanekar et al.}

\institute{National Centre for Radio Astrophysics, Post Bag 3, Ganeshkhind, Pune 411 007, India \and
Australia Telescope National Facility, CSIRO, Locked bag 194, Narrabri, 
NSW 2390, Australia \and
Institut d'Astrophysique de Paris - CNRS, 98bis Boulevard Arago, F-75014 Paris, France \and
Visiting Scientist, NFRA, P O Bus 2, 7990 AA Dwingeloo, The Netherlands.}

\date{Received mmddyy/ accepted mmddyy}
\offprints{Nissim Kanekar}

\maketitle

\begin{abstract}
\noi We report a deep search for 21~cm emission/absorption from the $z \sim 0.101$ 
candidate damped Lyman-$\alpha$ system towards PKS 0439$-$433, using the Australia 
Telescope Compact Array (ATCA). The spectrum shows a weak absorption feature --- at the 
$3.3 \sigma$ level --- which yields a lower limit of 730~K on the spin temperature of the 
system. No H~{\sc i} emission was detected: the $3\sigma$ upper limit on the H~{\sc i} mass of
the absorber is $2.25 \times 10^9 M_{\odot}$, for a velocity spread of $\sim 70$~km 
s$^{-1}$. The low H~{\sc i} mass and the high spin temperature seem to rule out 
the possibility that the absorber is a large gas-rich spiral galaxy.
\end{abstract}


\section{Introduction}
\label{sec:intro}

Damped Lyman-$\alpha$ (DLA) systems are objects with such high neutral 
hydrogen column densities ($N_{\rm H~{\sc i}} \ga 10^{20}$ cm$^{-2}$), that their 
optical depth in the damping wings of the Lyman-$\alpha$ line is 
appreciable. Given a  background quasar, this results in a broad 
absorption feature, easily detectable in even moderate resolution 
optical spectroscopy. DLA systems are the main repository of neutral
gas at high redshift ($z \sim 3$) and have, therefore, traditionally
been assumed to be the progenitors of large spiral galaxies (\cite{wolfe88,pw1,pw2}). 
In support of this hypothesis, the comoving mass density of neutral gas 
in these objects at $z \sim 3$ is comparable to the stellar mass density 
in bright galaxies today (\cite{sl2000,sl96,lanzetta91}), consistent with the 
gas having been converted into stars in the intervening period.

At low redshifts too, DLA systems are expected to be mainly associated
with spiral galaxies. Rao \& Briggs (1993) used the optical luminosity 
function and the average H~{\sc i} content of a given galactic morphological type
to conclude that the cross section for damped absorption at $z=0$ is dominated 
by large spiral galaxies; $\sim 90\%$ of the H~{\sc i} at $z=0$ resides in large spirals. 
An alternative to this hybrid approach to determining the local H~{\sc i} mass 
density is to directly use the observed H~{\sc i} mass function. The latter, as 
determined by blind H~{\sc i} surveys is, however, currently controversial.  
Zwaan~et~al.~(1997) find that the mass function is well fit by a Schecter 
function with a fairly flat $\alpha = -1.2$ slope at the faint end; this 
implies that the major contributors to the H~{\sc i} mass density at $z=0$ are 
$L^\star$ galaxies. On the other hand, Schneider~et~al.~(1998, see also 
Rosenberg~et~al.~2000) suggest that the space density of low mass 
(\MHI $< 10^8$ \Msun) galaxies is considerably larger than that 
predicted by the Schecter function fit of Zwaan~et~al.~(1997); a substantial 
fraction of the H~{\sc i} at $z = 0$ may then lie in smaller systems.

In recent times, the paradigm that damped absorption is predominantly 
associated with large disks has come under some scrutiny (\cite{hsr,ledoux,vladilo99}). 
Hubble Space Telescope (HST) and ground-based imaging of low and intermediate 
redshift damped systems (\cite{lebrun,rt98,raovla}) have shown that the absorbers 
are associated with a wide variety of galaxy types and not predominantly
with large spiral galaxies. In addition, DLA systems appear to have
low metallicities ($\sim 0.1$ solar) (\cite{pettini98}) and do not show the
$\alpha/$Fe enrichment pattern seen in low metallicity halo stars of the 
Milky Way (Centuri\'on et al. 2000; see, however, Lu et al. 1996). This suggests 
that the DLA systems have a different IMF and/or star formation history than large 
spirals. Finally, in cases where the background quasar is radio-loud, 21~cm absorption
studies have shown that the spin temperatures of the majority of DLA objects
are typically higher than those of local spirals (see Chengalur 
\& Kanekar (2000) and references therein). High spin temperatures arise 
naturally in dwarfs because these systems have low metallicities and pressures 
and, consequently, a larger fraction of the warm phase of H~{\sc i} as compared to 
large spiral disks (\cite{ck2000}). 

It is possible, of course, that the absorbing galaxies, though faint 
in the optical, are nonetheless exceedingly gas-rich, and their large H~{\sc i} 
envelopes cause them to be preferentially detected in absorption surveys.
At the very lowest redshifts, the latter hypothesis can be tested through
deep searches for H~{\sc i} 21~cm emission from the absorbers. Such searches yield
direct estimates of the H~{\sc i} mass and can thus be used to check whether or not 
the optically faint galaxies which give rise to DLA absorption have anomalously
large H~{\sc i} content. We describe, in this letter, a deep search for 
21~cm emission/absorption from a candidate damped absorber at $z \sim 0.101$
towards the quasar \pks (\cite{petitjean96}). The observations were carried 
out with the Australia Telescope Compact Array (ATCA).  No emission was 
detected, resulting in strong constraints on the H~{\sc i} mass of the absorber.

\section{Observations and Data analysis}
\label{sec:obs}

PKS~0439$-$433 was observed using the 1.5A configuration of the ATCA on 
a number of occasions in December~1999 and January~2000. The total on-source
integration time was $\sim65$~hours. A bandwidth of 8~MHz was used for 
the observations, divided into 1024 channels, and centred at 1290 ~MHz. 
This yielded a velocity resolution of $\sim$ 1.8 km s$^{-1}$ and a total 
velocity coverage of $\sim 1800$ km s$^{-1}$. Only XX and YY polarizations 
were measured. The strong source PKS~0438$-$436 was used for amplitude/phase and
bandpass calibration; this was observed every forty minutes in each 
observing session. 1934$-$638 was used as the primary flux calibrator, and 
observed at least once during each observing run.

The data were analyzed using the software package {\sc MIRIAD}.
Some baselines of the array were found to give intermittent correlator errors. 
Every 30-second averaged spectrum was hence inspected --- separately for each baseline --- 
to detect and reject any correlator errors as well as to reject any
obvious spectral interference. Next, the routine UVLIN was used to subtract 
out continuum emission from discrete sources in the field. 
A three-dimensional data cube was then made from the residual (i.e. continuum
subtracted) visibilities;  spectra for further analyses were constructed
using this cube. The data from the different observing sessions 
were shifted to the heliocentric frame before being averaged to 
produce the final spectrum.

While PKS~0439$-$433 is unresolved by the ATCA synthesized beam, the 
field also contains the much stronger compact source PKS~0438$-$436 (flux $\sim 
4.14$ Jy, our observations), near the edge of the primary beam, as well as one 
more weak compact source (flux $\sim$ 400 mJy, our observations). 
It was thus possible that the use of UVLIN for continuum subtraction might
result in spectral artefacts (\cite{cornwell}). 
We hence also tried an alternate analysis procedure to check for spectral errors 
that might arise due to the confusing sources in the field. This involved the 
use of the task UVSUB, to subtract out the individual point sources; UVLIN
was then run to remove any residual continuum emission and the 
resulting data set was again mapped to obtain the spectral data cube. The 
spectra obtained from the two procedures were found to be identical 
within the noise for each data set. 

\begin{figure}
\centering
\psfig{file=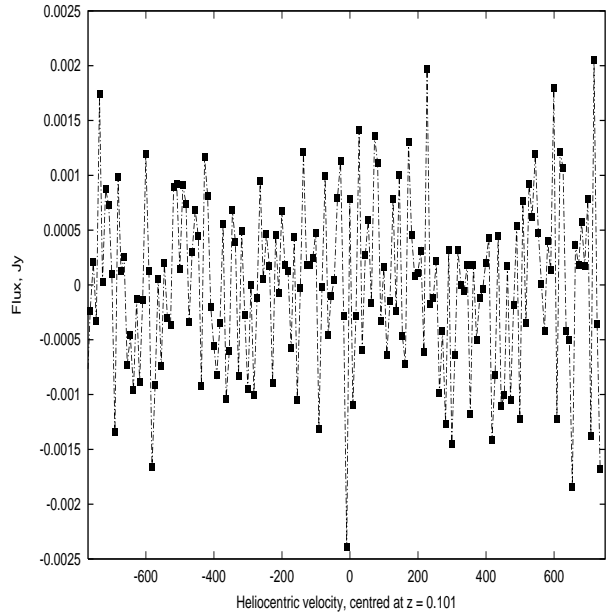,angle=-90,width=3.25truein,height=3.25truein}
\caption{ATCA 9 km s$^{-1}$ resolution spectrum towards \pks. 
The $x$-axis is heliocentric velocity in km s$^{-1}$, centred at
$z = 0.101$. Weak absorption can be seen, close to $v = 0$.}
\label{fig:fig2}
\end{figure}

\begin{figure}
\centering
\psfig{file=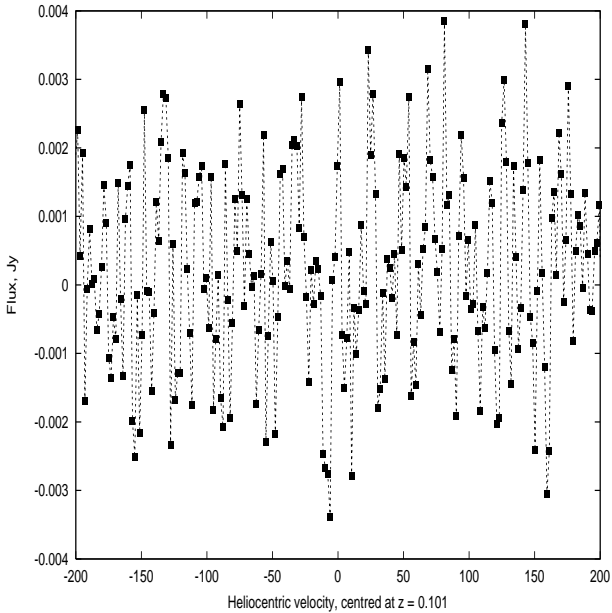,angle=-90,width=3.25truein,height=3.25truein}
\caption{Zoomed-in version of the ATCA 1.8 km s$^{-1}$ resolution 
spectrum towards \pks. The $x$-axis is heliocentric velocity in 
km s$^{-1}$, centred at $z = 0.101$.}
\label{fig:fig1}
\end{figure}

\begin{figure}
\centering
\psfig{file=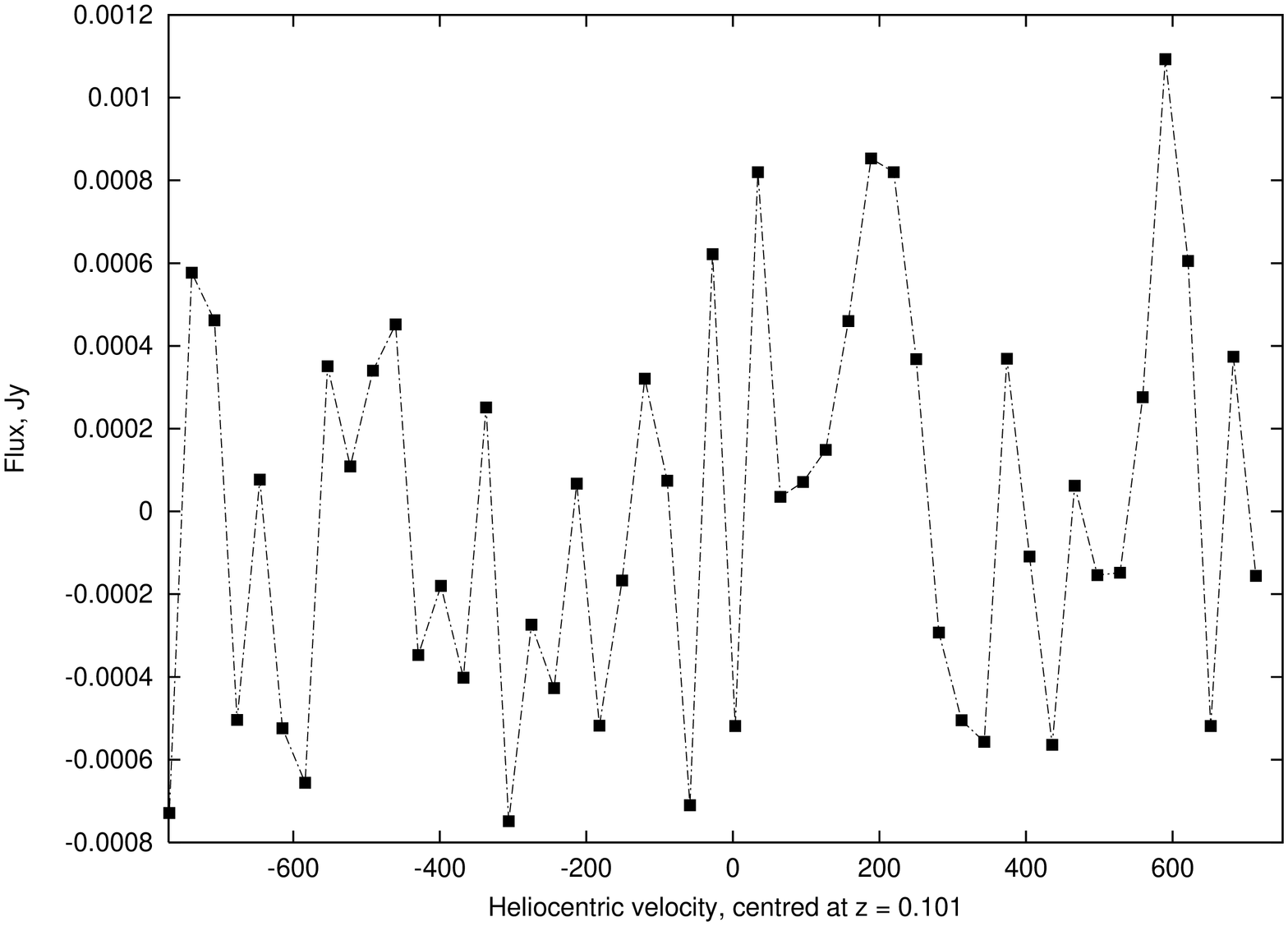,angle=-90,width=3.25truein,height=3.25truein}
\caption{ATCA 30 km s$^{-1}$ resolution spectrum towards \pks. The $x$-axis is
heliocentric velocity in km s$^{-1}$, centred at $z = 0.101$. This spectrum
was obtained after dropping antenna 6 and has a noise of 0.48 mJy.}
\label{fig:fig4}
\end{figure}

\begin{figure}
\centering\psfig{file=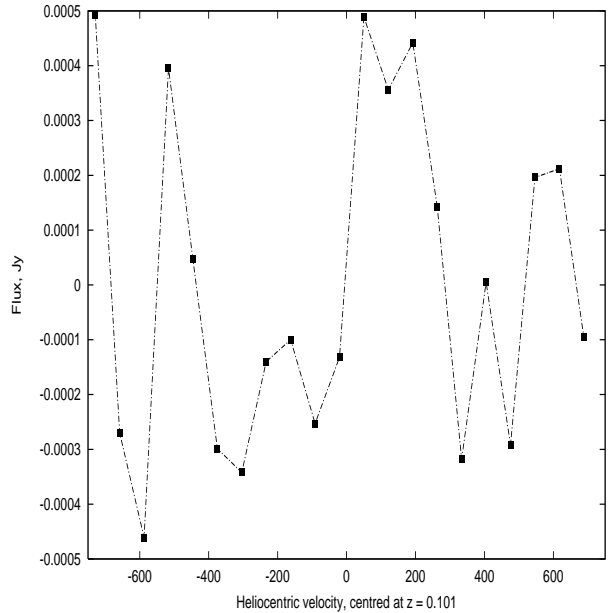,angle=-90,width=3.25truein,height=3.25truein}
\caption{ATCA 70 km s$^{-1}$ resolution spectrum towards \pks. The $x$-axis is 
heliocentric velocity in km s$^{-1}$, centred at $z = 0.101$. This spectrum
was also obtained after dropping antenna 6 and has a noise of 0.29 mJy.}
\label{fig:fig5}
\end{figure}

An RMS noise of 1.56~mJy was obtained per channel at the original 
velocity resolution of 1.8 km s$^{-1}$; this is close to the theoretical 
sensitivity of the ATCA, for our observing parameters. 
Figure~\ref{fig:fig2} shows the spectrum smoothed to a resolution of 
9 km s$^{-1}$: weak absorption  can be seen close to $z = 0.101$. This occurs 
at a heliocentric frequency of 1290.144 MHz, corresponding to a redshift of 
$0.10097 \pm 0.00003$; this is in good agreement with the redshift $z = 0.101$ 
obtained from metal lines (\cite{petitjean96}). The RMS noise on the spectrum is  
$\sim$ 0.76 mJy while the feature is $\sim$ 2.5 mJy deep (and only one channel wide), 
i.e. a 3.3$\sigma$ result. The absorption was seen in both the XX and 
YY polarizations separately, although, of course, at even lower significance 
levels. Figure~\ref{fig:fig1} shows a zoomed-in version of the spectrum 
at the original resolution of 1.8 km s$^{-1}$; the feature of figure 
\ref{fig:fig2} can be seen to be a few channels wide here (although 
within the noise). We note that the narrowness of the feature makes 
it very unlikely that it arises as an artefact from the continuum 
subtraction procedure. 

The search for H~{\sc i} emission was done after smoothing the spectrum to
a variety of velocity resolutions from 30 to 70 km s$^{-1}$; no 
statistically significant emission was seen at any resolution. The ATCA 
synthesized beam (in the 1.5A configuration), has a resolution of 
$\sim 15''$ on the sky, corresponding to a length scale of 18 kpc at 
the redshift of the absorber\footnote{An FRW cosmology with $\Omega = 1$ 
and $H_o = 75$ km s$^{-1}$ Mpc$^{-1}$ is used throughout this letter}. 
Since typical sizes of spirals tend to be larger than this 
($\sim 30 - 40$~kpc), there was a possibility that we might be resolving
out some of the emission. We therefore re-analyzed the data after flagging out 
antenna 6, which gives rise to the long ATCA baselines: this resulted 
in a synthesized beam of $\sim 40''$ on the sky, corresponding to a length 
scale of 50 kpc at $z = 0.101$. This spectrum was also searched for emission 
after smoothing to a variety of velocity resolutions between 30 km s$^{-1}$ 
and 70 km s$^{-1}$: no statistically significant emission was detected. The 
30 km s$^{-1}$ and 70 km s$^{-1}$ resolution spectra (after dropping antenna 6) 
are shown respectively in figures \ref{fig:fig4} and \ref{fig:fig5}; the 
RMS noise levels are 0.48 mJy and 0.29 mJy. Finally, the spectrum was also 
smoothed to a resolution of 200 km s$^{-1}$ to search for wide emission (note 
that this spectrum, which is not shown here, has only seven independent points). 
Again, no emission was seen; the RMS noise per 200 km s$^{-1}$ channel is 0.15 mJy.

\section{Discussion}
\label{sec:discuss}
\subsection{The spin temperature}
\noi For an optically thin, homogenous cloud, the 21~cm optical 
depth $\tau_{21}$ is related to the column density $N_{\rm H~{\sc i}}$ and the 
spin temperature $T_s$ by the expression 
(\cite{rohlfs86})
\begin{equation}
\label{eqn:Tspin}
        N_{\rm H~{\sc i}} = { 1.823\times10^{18} T_s \over f} \int \tau_{21} dV   \;\; ,
\end{equation}
\noi where $T_s$ is in K, $N_{\rm H~{\sc i}}$ in cm$^{-2}$ and dV in km s$^{-1}$. The 
covering factor $f$ can usually be estimated from VLBI observations;
unfortunately, no such observations exist in the literature for \pks. The
source, however, has an exceedingly flat spectrum, with flux values of 
330 mJy at 1.29 GHz (our observations), 320 mJy at 2.7~GHz and 300 mJy at
5~GHz (\cite{quiniento}), and is thus likely to be very compact. We hence 
assume a covering factor of unity; this should, of course, be verified by VLBI 
observations. Next, the $z = 0.101$ absorber towards PKS~0439$-$433 is 
a candidate damped system, with strong Mg~{\sc II}, Al~{\sc II}, Fe~{\sc II}, Si~{\sc II} 
and C~{\sc IV} absorption lines seen in the HST spectrum (\cite{petitjean96}).
The equivalent width ratios of these low-ionization lines indicate that the 
system is probably damped (see also Rao \& Turnshek 2000), with $N_{\rm H~{\sc i}}
\sim 10^{20}$~cm$^{-2}$ (although the Lyman-$\alpha$ line has itself not so far 
been observed for this absorber). This estimate of $N_{\rm H~{\sc i}}$ agrees 
well with the X-ray spectrum of \pks which shows absorption corresponding 
to $N_{\rm H~{\sc i}} = 2.3 \pm 0.8 \times 10^{20}$ per cm$^2$ (\cite{wilkes92}), 
with a Galactic contribution of $1.3 \pm 0.1 \times 10^{20}$ per cm$^2$
(\cite{lockman95}). The system is thus quite likely to be a moderate column
density damped absorber. Our 9 km s$^{-1}$ resolution ATCA spectrum has 
a peak optical depth $\tau_{max} \sim 0.0076$; equation (\ref{eqn:Tspin}) then 
yields a column density of $N_{\rm H~{\sc i}} = 1.4 \times 10^{17} \ T_s$ cm$^{-2}$. 
Combining this with the column density of Petitjean et al. (1996) gives a spin 
temperature of $\sim 730$ K. Note that this is a {\it lower} limit to the spin 
temperature, since the upper limit to the optical depth is $\sim 0.0076$.
If the $3.3 \sigma$ feature of figure \ref{fig:fig2} is {\it not} real,
it would imply that the system has an even higher spin temperature. 
As discussed in Chengalur \& Kanekar (2000), spiral galaxies tend to have 
low spin temperatures ($T_s \la 300$ K); it is thus unlikely that the absorber 
is a large spiral galaxy. 

\subsection{The H~{\sc i} mass of the absorber}

The non-detection of emission can be used to place limits on the 
H~{\sc i} mass of the  $z \sim 0.101$ absorber. Of course, the peak line flux 
is a function of the velocity distribution in the emitting gas; for a 
given H~{\sc i} mass, a cloud with a larger velocity dispersion will result in
a lower peak flux. We will, for simplicity, asssume a uniform velocity 
distribution with a width $\Delta V$. In this case, the H~{\sc i} mass of the 
emitting cloud is related to $\dV$ and the line flux $S$ by the 
expression 
\beq
\label{eqn:flux}
M_{\rm H~{\sc i}} = 2.35 \times 10^5 S \dV D_L^2 \; \; ,
\eeq
\noi where $\dV$ is in km s$^{-1}$, $S$ is in Jy and 
$M_{\rm H~{\sc i}}$ is in solar masses. $D_L$ 
is the luminosity distance of the cloud in Mpc ($D_L = 413.72$ Mpc, 
for a system at  $z = 0.101$). 

Velocity widths in dwarf galaxies are of the order of a few tens 
of km s$^{-1}$. On the other hand, in the case of spirals, the kinematics are
dominated by rotation, and the velocity spreads of H~{\sc i} emission profiles 
depend crucially on the inclination of the system to the line of sight. 
Face-on spirals tend to have velocity widths similar to those of dwarfs, 
$\dV \sim 20 - 30$ km s$^{-1}$; however, disks with large inclinations 
can have emission spread over a few hundred km s$^{-1}$. Even in the 
latter cases, however, velocity crowding in the tangent points results in a
characteristic `double-horned' profile, where each horn is 
$\sim 30 - 40$ km s$^{-1}$ wide, with a wide plateau in between. In the 
intermediate case, of inclination angles of the order of $45^\circ$, velocity 
widths of $\sim 70 - 100$ km s$^{-1}$ are common. We will hence use two 
velocity widths, $\dV = 30$ km s$^{-1}$ and $\dV = 70$ km s$^{-1}$, to obtain
limits on the H~{\sc i} mass of the absorber. 

The 30 km s$^{-1}$ resolution spectrum, with angular resolution $\sim 40 ''$ 
(i.e. without antenna~6), yields a $3\sigma$ upper limit of 1.44 mJy on the 
line flux. If we assume the emission profile has $\dV \sim 30$ km s$^{-1}$,
equation (\ref{eqn:flux}) gives $M_{\rm H~{\sc i}}(3\sigma) < 1.6 \times 10^9 
M_{\odot}$. The 70 km s$^{-1}$ resolution spectrum has a $3\sigma$ upper
limit of 0.87 mJy on the emission; for $\dV = 70$ km s$^{-1}$, this gives
$M_{\rm H~{\sc i}}(3\sigma) < 2.25 \times 10^9 M_{\odot}$. Both the above
estimates are smaller than the H~{\sc i} mass of the Milky Way, for which 
\MHI~$\approx 5 \times 10^9$~\Msun. Finally, the 200 km s$^{-1}$ resolution 
spectrum also rules out the possibility that the emission is spread over a 
wide velocity range; this spectrum yields an upper limit of $M_{\rm H~{\sc i}} 
(3\sigma) < 3.3 \times 10^9$~\Msun, again smaller than the ${\rm H~{\sc i}}$ 
mass of the Milky Way.

Petitjean et al. (1996) identified the absorber as an $L \sim 0.45 
L^\star$ spiral galaxy (impact parameter $\sim$ 7 kpc), from its colours in 
ground based images; they also estimated an inclination angle of 51$^\circ$, 
assuming a disk morphology. The spatial resolution of the imaging was, however, 
not high enough to confirm the morphology. Our observations, on the other hand,
 show that the ${\rm H~{\sc i}}$ content of the absorber is less than that of 
normal spirals like the Milky Way, and seem to rule out the possibility that 
it is a large, gas-rich spiral galaxy. It is, of course, possible that the 
system is an early-type, gas-poor spiral. High resolution HST imaging of the 
absorber is necessary to resolve this issue.

The $z \sim 0.101$ absorber towards \pks is the second low-redshift damped
Lyman-$\alpha$ system which has been searched for 21~cm emission, the other being the
$z = 0.0912$ absorber towards the quasar OI363 (\cite{em2000}). Both systems have 
high spin temperatures ($T_s \sim 775$ K for the $z = 0.0912$ system, 
Chengalur \& Kanekar 1999, Lane et al. 2000a), far higher than those of local
spirals or of damped systems which have been identified as spiral galaxies 
($T_s \sim 150 - 200$ K). The non-detection of emission in both cases (with 
fairly stringent limits on the H~{\sc i} mass) is consistent with the suggestion
of Chengalur \& Kanekar (2000) that the high $T_s$ damped absorbers are 
likely to be low-mass galaxies.

\noi {\bf Acknowledgements} It is a pleasure to thank Bob Sault for help in 
installing {\sc MIRIAD} as well as with analysis procedures. NK thanks the 
ATNF for hospitality during and after the observations. The Australia Telescope 
is funded by the Commonwealth of Australia for operation as a National Facility 
managed by CSIRO. This work was partly funded by a bezoekersbeurs from NWO
to JNC. JNC is grateful for the hospitality offered by NFRA for the period
during which part of this work was done.

\end{document}